\documentclass[conference]{IEEEtran}
\usepackage{graphicx,amsmath,amssymb}
\usepackage{subfigure}
\usepackage{citesort}
\usepackage{fancyhdr}
\usepackage{mdwmath}
\usepackage{mdwtab}
\usepackage{balance}
\usepackage{xcolor}
\usepackage{bm}
\usepackage{bbm}
\usepackage{stfloats}
\usepackage{cite}
\usepackage{epstopdf}
\usepackage{algorithm}
\usepackage{algorithmic}
\usepackage{multirow}
\usepackage{flafter}

\usepackage{amssymb}
\usepackage{amsmath}

\usepackage[
top    = 0.7 in,
bottom = 1.05 in,
left   = 0.63 in,
right  = 0.63 in]{geometry}

\newtheorem{theorem}{Theorem}

\newtheorem{lemma}{Lemma}

\newtheorem{corollary}{Corollary}

\begin{document}
\title{Multi-cell NOMA: Coherent Reconfigurable Intelligent Surfaces Model With Stochastic Geometry}

\author{
\IEEEauthorblockN{Chao~Zhang\IEEEauthorrefmark{1}, Wenqiang~Yi\IEEEauthorrefmark{1}, Yuanwei~Liu\IEEEauthorrefmark{1}, Qiang~Wang\IEEEauthorrefmark{2}}
\IEEEauthorblockA{\IEEEauthorrefmark{1}Queen Mary University of London, London, UK}
\IEEEauthorblockA{\IEEEauthorrefmark{2}Beijing University of Posts and Telecommunications, Beijing, China} }

\maketitle

\begin{abstract}

$\Theta$ $\theta$
 Reconfigurable intelligent surfaces (RISs) become promising for enhancing non-orthogonal multiple access (NOMA) systems, i.e., enhancing the channel quality and altering the SIC orders. Invoked by stochastic geometry methods, we investigate the downlink coverage performance of RIS-aided multi-cell NOMA networks. We first derive the RIS-aided channel model, concluding the direct and reflecting links. The analytical results demonstrate that the RIS-aided channel model can be closely modeled as a Gamma distribution. Additionally, interference from other cells is analyzed. Lastly, we derive closed-form coverage probability expressions for the paired NOMA users. Numerical results indicate that 1) although the interference from other cells is enhanced via the RISs, the performance of the RIS-aided user still enhances since the channel quality is strengthened more obviously; and 2) the SIC order can be altered by employing the RISs since the RISs improve the channel quality of the aided user.
\end{abstract}

\vspace{-0.1cm}
\section{Introduction}

As a promising technique, non-orthogonal multiple access (NOMA) achieves new freedom by sharing spectrum with power or code multiplexing methods. With the aid of successive interference cancellation (SIC), various users in the same resource block can be split and decoded, thereby the user connectivity and the spectrum efficiency are enhanced significantly~\cite{5G}. In spite of benefits, several implementation challenges occur in the following aspects. 1) On the one hand, the received power of the final decoded user is low while the interference is increased from other NOMA users. In general, far users will be decoded at the last decoding order, thereby the NOMA technique may contribute to the atrocious performance of the far users. 2) On the other hand, the fixed decoding order based on channel quality may cause a waste of resources. We assume the first decoded user can achieve $10$ BPCU (bit per cell use) but this user only need $6$ BPCU to transmit the message, which leads to $4$ BPCU wasted. In order to cope with the challenges, reconfigurable intelligent surfaces (RIS) are taken into the consideration.

The material of a RIS is considered as a two-dimensional-equivalent reconfigurable meta-material, in which several elements such as scattering particles or meta-atoms are involved \cite{RISaccess}. With the aid of meta-surfaces, the reflecting direction can be altered by the RIS towards the targeted user to obtain the perfect electric field strength. Hence, with coherent reflecting links, the RISs can enhance the channel quality of its aided user. By exploiting the RISs near the last decoded user (the far user), the far user's channel gain is improved, thereby 1) the far user's received power is significantly enhanced; and 2) the decoding order can be changed to utilize the resources perfectly. In this model, we harness stochastic geometry methods to evaluate the spacial effects of multi-cell networks \cite{nNearestUser}. With the aid of the RISs and NOMA technique, the motivations and contributions are revealed in the following.

Motivated by 1) enhancing the channel quality of the last decoded NOMA user; and 2) altering the SIC order to maintain that the user who required a high transmit rate is employed at the first stage of the SIC order, we investigate RIS-aided multi-cell NOMA networks. The main contributions can be summarized as: 1) we first derive the RIS-aided channel model including the direct and reflecting links, which reveals that the channel model in power domain can be expressed as a Gamma distribution; 2) we derive closed-form expressions of coverage probability for the paired NOMA users; and 3) numerical results illustrate that both of the interference from other cells and the channel quality of the RIS-aided user enhance but the enhancement of the channel quality dominates the upper hand.

\vspace{-0.1cm}
\section{System Model}

 \begin{figure*}[t]
 \vspace{-0.2cm}
\centering
\includegraphics[width= 6.5in]{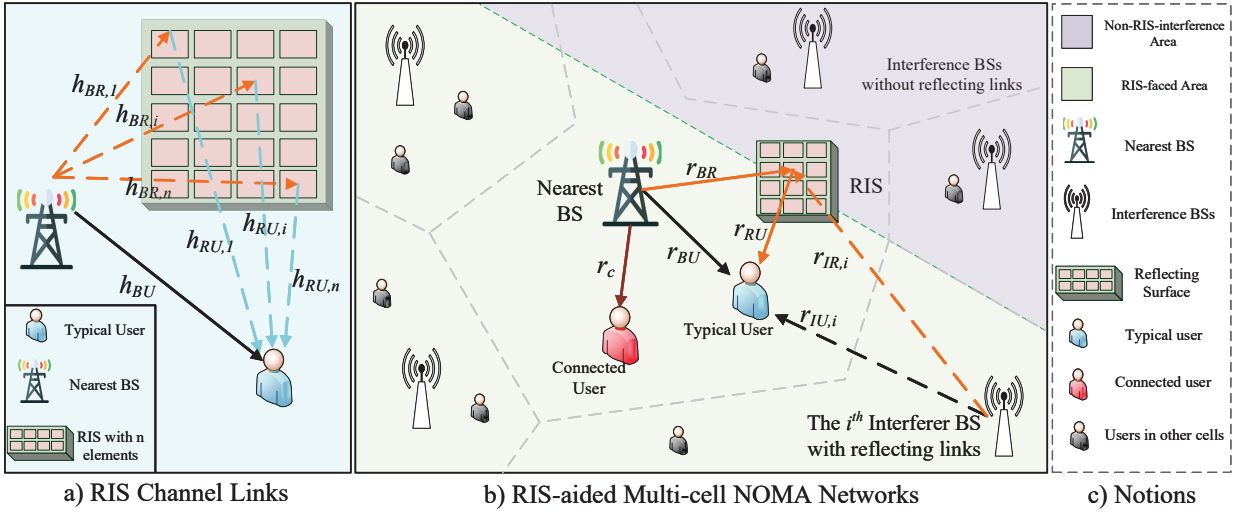}
\caption{Indication of the signal model: (a) RIS-aided channel links with $n$ elements: the direct and reflecting links; (b) RIS-aided multi-cell NOMA networks: a NOMA pair concluding a typical user and a connected user as a treatise; (c) Notions.}
\label{fig_1}
\vspace{-0.2cm}
\end{figure*}

In this paper, RIS-aided downlink NOMA multi-cell networks are taken into consideration, which concludes randomly deployed base stations (BSs), users and RISs. With the respect to NOMA clusters, we evaluate a two-user scenario in this treatise, where different NOMA pairs are served by orthogonal frequencies to cancel intra-cell interference. We define one of the deployed users as the typical user, which is exploited in a typical NOMA pair. Additionally, we assume that the other paired user in the same resource block has already been connected to the targeted BS. An RIS is allocated near the typical user to enhance the performance of the users.

\subsection{Deployment of Devices}

Invoked by stochastic geometry models, we deploy BSs and users as two independent homogeneous Poisson point process (HPPP): 1) $\Phi_u\subset \mathbb{R}^2$ with density $\lambda_u$; and 2) $\Phi_b\subset \mathbb{R}^2$ with density $\lambda_b$. Among these users, a typical user is randomly chosen from $\Phi_u$ and its location is fixed at the origin of the considered plane $O$. For the typical user network, it is deployed by a simplified Matern cluster process (MCP) pattern, which is one of Poisson cluster process (PCP) models with a fixed number of nodes and fixed positions in each cluster \cite{8856258}. More specifically, this pattern concludes two processes, i.e., a parent process to deploy the clusters and a daughter process to allocate the nodes in each cluster. In this typical user network, we consider that the deployment of BSs as the parent process is an independent HPPP, which has been denoted as $\Phi_b\subset \mathbb{R}^2$ with density $\lambda_b$. Additionally, we simplify the daughter process with a fixed distance between the typical user and its aided RIS, namely $r_{RU}$. To simplify the analysis, the connected user is not included in the user set $\Phi_u$ and the distance between this user to its BS is invariable as $r_c$.

The association criterion for the typical user is to select the BS with the highest received power, which means that the distance between the RIS and the associated BS is the nearest. Based on the typical user as the origin, the locations of the RIS and the associated BS are denoted as $r_{RU} = \mathbf{x}_R$ and $r_{BU} = \mathbf{x}_{B,near}$. Therefore, the distance between the associated BS and the RIS is correspondingly expressed as
 \begin{align}
 r_{BR} = \mathbf{x}_{BR} = \arg \min_{\mathbf{x}_B \in \Phi_{B}} \|\mathbf{x}_B-\mathbf{x}_R \|
 \end{align}
 where $r_{RU} = \mathbf{x}_R$ is fixed, $\Phi_{B} \subset\Phi_b$, $\Phi_{B} \subset \mathbb{O}(R_L,\infty)$\footnote{The $\mathbb{O}(a,b)$ represents an annulus with the inner radius $a$ and outer radius $b$.} and the location of arbitrary interfering BS can be denoted by $\mathbf{x}_I \in \Phi_{b} \setminus  \mathbf{x}_{B,near}$.

We assume the RIS is deployed near the typical user, thereby the distances $r_{BR}$ and $r_{BU}$ are approximately equal, denoted as $r_{BR} \approx  r_{BU}$. Thus, based on 2-D HPPP statistics \cite{nNearestUser, nNearestUserJ}, the PDF of the distance between a BS and its $n^{th}$ nearest BS is derived as
\begin{align}\label{r_BU_n}
{f_{{r_{BU}}}}\left( {x,n} \right) = \frac{{2{{\left( {\pi {\lambda _b}} \right)}^n}}}{{\left( {n - 1} \right)!}}{x^{2n - 1}}\exp \left( { - \pi {\lambda _b}{x^2}} \right),
\end{align}
thereby the PDF of $r_{BR}$ and $r_{BU}$ can be derived by \eqref{r_BU_n} with $n=1$ as
\begin{align}\label{r_BU}
{f_{{r_{BR}}}}\left( {x} \right)={f_{{r_{BU}}}}\left( {x} \right) = 2\pi {\lambda _b}{x}\exp \left( { - \pi {\lambda _b}{x^2}} \right).
\end{align}

\subsection{RIS-aided Channel Model}

We assume the RIS has $n$ elements, and the BSs and the users are equipped with a single antenna. Based on the deployment of the typical network, there are three categories of important communication links in the considered NOMA group: 1) BU links, the links between the typical user and the nearest BS; 2) RU link, the link between the RIS and the typical user; and 3) BR link, the link between the BS and the employed RIS. The BU links are the direct links for the typical user. Additionally, the BR and RU links are compositions of reflecting links aided by the employed RIS. Since the RIS is made of elementary elements called scattering particles or meta-atoms, which is capable of altering the wavefront of the radio waves, the phase responses for the RIS reflect links and direct links can be altered as the same. Hence, with the aid of RISs, the performance of the typical enables to be enhanced significantly.

\subsubsection{Multi-path Model}
We assume all the links of the typical and connected users are Rayleigh fading channels, i.e., $h_{BU}$ for BU links, $h_{RU,i}$ for RU links, $h_{BR,i}$ for BR links and $h_{c}$ for the links between the connected user and its BS, where $i \in \left[1,n\right]$. Hence, the probability density function (PDF) and cumulative distribution function (CDF) for Rayleigh fading channels can be expressed as $f(x) = 2x\exp \left( { - {x^2}} \right)$ and $F(x) = 1-\exp(-x^2)$. Since it is assumed that the RIS is employed near the typical user, we can assume the channel gain of RU links is a constant.

\subsubsection{Interference Analysis}
Since the RIS only reflects the signals faced the surface, the interference BSs are split into two portions: 1) the interference BSs aided by the employed RIS, and 2) the interference BSs without RIS. This is the key difference between RIS-aided multi-cell networks and conventional multi-cell networks. We define a coefficient $\rho_I \in [0,1]$ to indicate the percentage of the interference BSs with the aid of RISs, thereby the rest BSs without the aid of RISs is $1-\rho_I$.

\subsubsection{Path Loss Model}

For each NOMA pair, the path loss model for the three links can be defined as the conventional wireless communication models. Therefore, the path loss expressions of the connected user and the typical user are as
 \begin{align}
 &P_{BR}(\mathbf{x}_{B,near},\mathbf{x}_R)=C_{BR}\|\mathbf{x}_{B,near}-\mathbf{x}_R\|^{-\alpha_t},\\
  &P_{RU}(O,\mathbf{x}_R) = C_{RU}\|\mathbf{x}_R\|^{-\alpha_t},\\
 & P_{BU}(O,\mathbf{x}_{B,near})= C_{BU}\|\mathbf{x}_{B,near}\|^{-\alpha_t},\\
  &P_c(\mathbf{x}_{B,near},\mathbf{x}_C) = C_c\|\mathbf{x}_{B,near}-\mathbf{x}_C\|^{-\alpha_c},
 \end{align}
where the $ \{C_{BR},C_{RU},C_{BU},C_c\}$ are the intercepts for different links, $\alpha_c$ is the path loss exponent for the connected user and $\alpha_t$ is the path loss exponent for the typical user (including the direct and reflecting links). Note that the distance between the connected user and the associated BS is fixed. Therefore, $r_c = \|\mathbf{x}_{B,near}-\mathbf{x}_C\|$ is a constant. Since we assume the RIS is employed near the typical user, we assume the channel gain is constant, denoted as $\left|h_{RU,i}\right|^2P_{RU}(O,\mathbf{x}_R) =A$.

\subsection{Signal Model}
We assume the connected user as the near user and the typical user as the far user, thereby we exploit RISs to enhance the performance of the far user. In the general NOMA case, the near user will accomplish the SIC process and the messages from the far user are decoded firstly. However, since we harness the RISs to enhance the channel quality of the far user, the rank of the channel quality of two paired NOMA users can be altered by adjusting the number of RIS's elements $n$. Thus, we enable to acquire high flexibility on SIC decoding orders with various values of $n$. In this treatise, we consider the scenario where the strength of channel quality for the far user is better than that of the near user since $n$ is large enough to change the SIC order. Hence, the far user will cope with the SIC order and the messages of the near user are at the first stage of the SIC process.

In order to guarantee the quality of service (QoS) of NOMA users, we assume more transmit power is allocated from the associated BS to the typical user than the connected user. Therefore, the signal-to-noise-ratio (SINR) of the SIC process at the typical user is expressed as
  \begin{align}
{\gamma _{{\rm{SIC}}}} = \frac{{{a_c}{P_t}{{\left| {\bf{{G_{RU}}}\Theta }{\bf{G_{BR}}} + {G_{BU}} \right|}^2}}}{{{a_t}{P_t}{{\left| {\bf{{G_{RU}}}\Theta }{\bf{G_{BR}}} + {G_{BU}} \right|}^2} + {I_{t,{\rho _I}}} + {\sigma ^2}}},
 \end{align}
 where
 \begin{align}
{I_{t,{\rho _I}}}& = {\rho _I}\sum\limits_{{{\bf{x}}_I} \in {\Phi _b} \setminus {{\bf{x}}_B}} {{P_t}{{\left| {{{\bf{G}}_{{\bf{RU}}}}{\bf{\Theta }}{{\bf{G}}_{{\bf{BR}}}} + {G_{BU}}} \right|}^2}} \notag\\
&\hspace*{0.3cm}{\rm{ + }}\left( {1 - {\rho _I}} \right)\sum\limits_{{{\bf{x}}_I} \in {\Phi _b} \setminus {{\bf{x}}_B}} {{P_t}{{\left| {{h_{BU}}} \right|}^2}{P_{BU}}(O,{{\bf{x}}_{B,near}})},
 \end{align}
 and $a_c$ and $a_t$ are the power allocation parameters for the connected and the typical user, $P_t$ is the transmit power of BSs in NOMA clusters and $\sigma ^2$ is the variance of additive white Gaussian noise (AWGN). With respect to the channel matrixes, ${G_{BU}} = {h_{BU}}\sqrt {{P_{BU}}(O,{{\bf{x}}_{B,near}})} $, ${\mathbf{G_{RU}}} = {\mathbf{H_{RU}}}\sqrt {{P_{RU}}(O,{{\bf{x}}_R})} $, $\mathbf{{G_{BR}}} = {\mathbf{H_{BR}}}\sqrt {{P_{BR}}({{\bf{x}}_{B,near}},{{\bf{x}}_R})} $, $\mathbf{H_{BR}} \buildrel \Delta \over = [h_{BR,1},h_{BR,2},\cdots,h_{BR,n}]^H$ is the $n \times 1$ channel gain matrix of $BR$ links, $\mathbf{H_{RU}}  \buildrel \Delta \over = [h_{RU,1},h_{RU,2},\cdots,h_{RU,n}]$ is the $1 \times n$ channel gain matrix of $RU$ links, ${\bf{\Theta }} \buildrel \Delta \over = diag[{\beta _1}{\phi _1},{\beta _2}{\phi _2}, \cdots ,{\beta _n}{\phi _n}]$ is the diagonal matrix, where $\beta_n \in [0,1]$ represents the power consumption coefficient of RISs and $\phi_n =\exp(j\theta_n)$ with $j=\sqrt{(-1)}$ and $\theta_n \in [0,2\pi)$. Additionally, we set $a_c > a_t$ and $a_c + a_t = 1$. Regarding the interference $I_{t,\rho_I}$, we assume $\rho_I \in [0,1]$ to evaluate the amount of BSs with RIS-aided links since the signal from the back of RISs cannot pass through the RISs, thereby the interference from other cells is split into two parts.

Due to the SIC process, the messages of the connected user can be removed, followed by decoding the data for the typical user. The decoding SINR after the SIC process for the typical user can be expressed as
\begin{align}
{\gamma _t} = \frac{{{a_t}{P_t}{{\left| {{\mathbf{G_{RU}}}\mathbf{\Theta {G_{BR}}} + {G_{BU}}} \right|}^2}}}{{{I_{t,{\rho _I}}} + {\sigma ^2}}}.
\end{align}

The connected user enables to directly decode its messages by regarding the other NOMA user's signal as interference. Hence, we can express the decoding SINR for the typical user as
\begin{align}
{\gamma _{\rm{c}}} = \frac{{{a_c}{P_t}{{\left| {{h_c}} \right|}^2}{P_c}({{\bf{x}}_{B,near}},{{\bf{x}}_C})}}{{{a_t}{P_t}{{\left| {{h_c}} \right|}^2}{P_c}({{\bf{x}}_{B,near}},{{\bf{x}}_C}) + {I_c} + {\sigma ^2}}},
 \end{align}
 where
  \begin{align}
{I_c} = \sum\limits_{{{\bf{x}}_I} \in {\Phi _b} \setminus {{\bf{x}}_B}} {{P_t}{{\left| {{h_c}} \right|}^2}{P_c}({{\bf{x}}_I},{{\bf{x}}_C})} .
 \end{align}

\section{RIS-aided Channel Model}

The channels between the typical user and its associated BS are split into two portions: 1) the reflecting links via RIS; and 2) the direct link via conventional wireless communication methods. When signals with the same frequency and waveform are transmitted from the BS, we consider the signals from various links, i.e., the reflecting and direct links, are coherent. In general, when we design the passive beamforming at the RISs, two directions are involved: 1) one to obtain the maximum received power; and 2) the other to reduce the interference via signal cancellation \cite{yuanwei_mag}. In this treatise, we aim to enhance the performance of the typical user by the RISs, thereby we assume that a perfectly coherent scenario is achieved by passive beamforming, which means all channels are adjusted with the same phase.

We denote an equivalent channel between the typical user and its BS as $g_{BU}  \buildrel \Delta \over = {\mathbf{g_{RU}}}\mathbf{\Theta {G_{BR}}} +$ $ {G_{BU}} $. We assume $h_{BR,i} = {c_{BR,i}}\exp \left( { - j{\theta _{BR,i}}} \right)$ with $i \in [1,n]$, and $h_{BU} = {c_{BU}}\exp \left( { - j{\theta _{BR,i}}} \right)$, which are Rayleigh variables in complex form. With the aforementioned assumption that the RISs and the typical user are deployed near, we can simplify the equivalent channel model $g_{BU}$ as
\begin{align}
{g_{BU}} &= \sqrt {{C_t}d_t^{ - {\alpha _t}}} \left( {A\sum\limits_{i = 1}^n {{\beta _i}{c_{BR,i}}} \exp \left( { - j\left( {{\vartheta _i} + {\theta _{BR,i}}} \right)} \right)} \right.\notag\\
&\hspace*{0.3cm}\left. { + {c_{BU}}\exp \left( { - j{\theta _{BU}}} \right)} \right)
=\sqrt {{C_t}d_t^{ - {\alpha _t}}} f_{BU},
\end{align}
where $f_{BU}$ is the small scale fading for the RIS-aided channel, $P_{BR}(\mathbf{x}_{B,near} )= P_{BU}(O,\mathbf{x}_{B,near})={C_t}d_t^{ - \alpha_t }$, $C_t$ is the intercept and $d_t$ is the distance between the typical user and its BS, which has the PDF as \eqref{r_BU}.

Based on the assumption of the perfectly coherent scenario, the phases of all the links can be adjusted as the result of $ {\vartheta _i} + {\theta _{BR,i}} = {\theta _{BU}} = \theta $ for all $i \in [1,n]$. Hence, we can express the equivalent channel model in power domain as
\begin{align}
 {\left| {{g_{BU}}} \right|^2} = {C_t}d_t^{ - {\alpha _t}}{\left( {\sum\limits_{i = 1}^n {A{\beta _i}{c_{BR,i}}}  + {c_{BU}}} \right)^2}.
\end{align}

\begin{theorem}\label{Channel1}
Since we consider the sub-links, i.e., the reflecting links through each RIS element and the direct link, are independent Rayleigh fading channels, $c_{BU}$ and $c_{BR,i}$ (for $i \in [1,n]$) are $n+1$ independent and identically distributed variables. We assume the power consumption coefficients for all RIS elements are the same, denoted as $\beta_1 = \beta_2=\cdots =\beta_n=\beta$. Hence, the distribution of the channel model ${\left| {{g_{BU}}} \right|^2}$ in power domain is derived as
\begin{align}\label{channelintegral}
{f_{{{\left| {{g_{BU}}} \right|}^2}}}\left( x \right) = \frac{1}{{2\Lambda \sqrt {\frac{x}{\Lambda }} }}{\cal L}_{{S_k}}^{ - 1}\left\{ {{{\left( {\frac{1}{2}\Psi \left( {1,\frac{1}{2};\frac{{{s^2}}}{4}} \right)} \right)}^K}} \right\}\left( {\sqrt {\frac{x}{\Lambda }} } \right),
\end{align}
where $\Lambda = {C_t}d_t^{ - {\alpha _t}}{\left( {A\beta } \right)^2}$, $K = n+1$, ${\cal L}_{{S_k}}^{ - 1}\left\{ \cdot \right\}$ is inverse Laplace transform, $\Psi \left( {\cdot,\cdot;\cdot} \right)$ is Tricomi's confluent hypergeometric function and $\Psi \left( {1,\frac{1}{2};z} \right) = 2 - 2\exp \left( z \right)\sqrt \pi  \sqrt z {\rm{erfc}}\left( {\sqrt z } \right)$ is a special case for Tricomi's function.
\begin{IEEEproof}
See Appendix~A.
\end{IEEEproof}
\end{theorem}

Since the inverse Laplace transform of ${{{\left( {\frac{1}{2}\Psi \left( {1,\frac{1}{2};\frac{{{s^2}}}{4}} \right)} \right)}^K}}$ is tough to be obtained in \textbf{Theorem \ref{Channel1}}, we cannot derive an efficient and concise distribution. Based on the simulation results, the trend of the CDF for the channel distribution in power domain can be closely approximated to a Gamma distribution. Hence, we exploit a Gamma distribution to fit the targeted RIS channel as \textbf{Corollary \ref{nihe}} \cite{Refnihe}.

\begin{corollary}\label{nihe}
The exact distribution for the RIS-aided channel model including the direct and reflecting links can be fitted as a Gamma distribution. With the aid of Matlab curve fitting tools, the PDF and CDF for the small-scale fading model ${\left| {{f_{BU}}} \right|^2}$ in power domain can be expressed as
\begin{align}\label{EQ_nihe}
{f_{{{\left| {{f_{BU}}} \right|}^2}}}\left( x \right) &= \frac{{{x^{a - 1}}}}{{\Gamma \left( a \right){b^a}}}\exp \left( { - \frac{x}{b}} \right),\\
{F_{{{\left| {{f_{BU}}} \right|}^2}}}\left( x \right) &= \frac{{\gamma \left( {a,x/b} \right)}}{{\Gamma \left( a \right)}},
\end{align}
where $a$ and $b$ are curve fitting coefficients with $a = n$, the scale coefficient $b\approx n $ when $\beta = 1$, $\Gamma(\cdot)$ is complete gamma function and $\gamma(\cdot,\cdot)$ is the lower incomplete gamma function, denoted as $\gamma(s,x)=\int_0^x t^{s-1}e^{-t}dt$.
\end{corollary}

\section{Performance Evaluation}

We evaluate the coverage performance of the typical user and the connected user in this section. The expression of coverage probability for the two users can be expressed as
\begin{align}\label{EQ_nihe}
&{{\rm{P}}_{cov,t}} = \Pr \left\{ {{\gamma _{{\rm{SIC}}}} > \gamma _{{\rm{SIC}}}^{th},{\gamma _t} > \gamma _t^{th}} \right\},\\
&{{\rm{P}}_{cov,c}} = \Pr \left\{ {{\gamma _{\rm{c}}} > \gamma _{\rm{c}}^{th}} \right\}\, ,
\end{align}
where $\gamma _{{\rm{SIC}}}^{th}$, $\gamma _t^{th}$ and $\gamma _{\rm{c}}^{th}$ are the coverage thresholds.

\subsection{Interference Analysis}
Before the performance analysis, the Laplace transforms of the typical and connected user are derived via \textbf{Lemma \ref{interference1}} and \textbf{Lemma \ref{interference2}}. For the typical user, we assume only the interference BSs that is facing the RISs have the RIS-aided channels, thereby the interference BSs are split into two portions, i.e., 1) BSs facing the RISs with RIS-aided channels; and 2) BSs behind the RISs with conventional wireless communication channel (Rayleigh fading channels). For the connected user, the interference BSs experience Rayleigh fading channels without the aid of RISs.

\begin{lemma}\label{interference1}
For the typical user, the Laplace transform of interference, concluding two portions of interference BSs, is derived as
\begin{align}\label{Lap_inter_1}
{\cal L}_{(s)}^t{\rm{ = }}\exp \left( { - \pi {\lambda _b}d_t^2\left( {{}_2{F_1}\left( { - \frac{2}{{{\alpha _t}}},1;1 - \frac{2}{{{\alpha _t}}}; - {\xi _{\rm{1}}}s} \right) - 1} \right)} \right)\notag\\
\times\exp \left( { - \pi {\lambda _b}d_t^2\left( {{}_2{F_1}\left( { - \frac{2}{{{\alpha _t}}},a;1 - \frac{2}{{{\alpha _t}}}; - {\xi _2}s} \right) - 1} \right)} \right),
\end{align}
where ${\xi _{\rm{1}}}{\rm{ = }}\frac{{\left( {1 - {\rho _I}} \right){P_t}{C_t}}}{{d_t^{{\alpha _t}}}}$, ${\xi _2} = \frac{{b{\rho _I}{P_t}{C_t}}}{{d_t^{{\alpha _t}}}}$ and ${}_2{F_1}\left( { \cdot , \cdot ; \cdot ; \cdot } \right)$ is the hypergeometric function.
\begin{IEEEproof}
See Appendix~B.
\end{IEEEproof}
\end{lemma}

\begin{lemma}\label{interference2}
For the connected user with a coefficient, defined as ${\xi _3} = \frac{{s{P_t}{C_c}}}{{r_c^{{\alpha _c}}}}$, the Laplace transform of interference can be derived as
\begin{align}\label{Lap_inter_2}
{\cal L}_{(s)}^c{ =} \exp \left( { - \pi {\lambda _b}r_c^2\left( {{}_2{F_1}\left( { - \frac{2}{{{\alpha _c}}},1;1 - \frac{2}{{{\alpha _c}}}; - {\xi _3}s} \right) - 1} \right)} \right),
\end{align}
\begin{IEEEproof}
See \textbf{Lemma \ref{interference1}}.
\end{IEEEproof}
\end{lemma}

\subsection{Coverage Probability}

Based on the derivations of the Laplace transform, we derive the coverage probability expressions of the typical user and connected user in the following.

\begin{theorem}\label{CPt}
Since the RISs enhance the channel quality of the typical user, we arrange the typical user complete the SIC procedure. Hence, the coverage probability concluding SIC process is derived as
\begin{align}\label{eqCPt}
{{\rm{P}}_{cov,t}} = & 2\pi {\lambda _b}\sum\limits_{k = 1}^a {{{\left( { - 1} \right)}^{k + 1}}\binom{a}{k}}I_1,
\end{align}
where $I_1 = \int_0^\infty  {x\exp \left( { - {\Xi _2}{x^{{\alpha _t}}}} \right)\exp \left( { - {\Xi _1}{x^2}} \right)} dx$, $\Upsilon  = \max \left( {\frac{{\gamma _{{\rm{SIC}}}^{th}}}{{\left( {{a_c} - \gamma _{{\rm{SIC}}}^{th}{a_t}} \right)}},\frac{{\gamma _t^{th}}}{{{a_t}}}} \right)$, ${\Xi _2} = \frac{{k{\eta _t}\Upsilon {\sigma ^2}}}{{{P_t}{C_t}}}$ and ${\Xi _1}$ is expressed as
\begin{align}
{\Xi _1} &= \pi {\lambda _b}{}_2{F_1}\left( { - \frac{2}{{{\alpha _t}}},1;1 - \frac{2}{{{\alpha _t}}}; - k{\eta _t}\Upsilon \left( {1 - {\rho _I}} \right)} \right)\notag\\
 &+ \pi {\lambda _b}\left( {{}_2{F_1}\left( { - \frac{2}{{{\alpha _t}}},a;1 - \frac{2}{{{\alpha _t}}}; - k{\eta _t}b\Upsilon {\rho _I}} \right) - 1} \right).
\end{align}
\begin{IEEEproof}
 After we defined a coefficient, denoted as $\Upsilon  = \max \left( {\frac{{\gamma _{{\rm{SIC}}}^{th}}}{{\left( {{a_c} - \gamma _{{\rm{SIC}}}^{th}{a_t}} \right)}},\frac{{\gamma _t^{th}}}{{{a_t}}}} \right)$, the coverage probability is expressed as ${{\rm{P}}_{cov,t}} = \Pr \left\{ {{{\left| {{f_{BU}}} \right|}^2} > {{\Upsilon d_t^{{\alpha _t}}\left( {{I_{t,{\rho _I}}} + {\sigma ^2}} \right)} \mathord{\left/
 {\vphantom {{\Upsilon d_t^{{\alpha _t}}\left( {{I_{t,{\rho _I}}} + {\sigma ^2}} \right)} {{P_t}{C_t}}}} \right.
 \kern-\nulldelimiterspace} {{P_t}{C_t}}}} \right\}$. with the aid of the closed boundary for Gamma distribution, i.e., ${{\rm{P}}_{{{\left| {{f_{BU}}} \right|}^2}}}\left\{ {{{\left| {{f_{BU}}} \right|}^2} < x} \right\} = {\left( {1 - \exp ( - {\eta _t}x)} \right)^N}$ with ${\eta _t} = \frac{{\rm{1}}}{b}{\left( {a!} \right)^{ - \frac{1}{a}}}$, the expression of the coverage probability is derived as
\begin{align}
{{\rm{P}}_{cov,t}} {=} \sum\limits_{k = 1}^a {{{\left( { - 1} \right)}^{k + 1}}\binom{a}{k}} {\rm{E}}\left[ {{e^{ - \frac{{k{\eta _t}\Upsilon d_t^{{\alpha _t}}}}{{{P_t}{C_t}}}{I_{t,{\rho _I}}}}}} \right]{\rm{E}}\left[ {{e^{ - \frac{{k{\eta _t}\Upsilon d_t^{{\alpha _t}}{\sigma ^2}}}{{{P_t}{C_t}}}}}} \right]\notag\\
{=}  \sum\limits_{n = 1}^a {{{\left( { - 1} \right)}^{k + 1}}}\binom{a}{k} {\rm{E}}\left[ {{e^{ - \frac{{k{\eta _t}\Upsilon d_t^{{\alpha _t}}{\sigma ^2}}}{{{P_t}{C_t}}}}}} \right]{\cal L}_{(s)}^t\left( {\frac{{k{\eta _t}\Upsilon d_t^{{\alpha _t}}}}{{{P_t}{C_t}}}{I_{t,{\rho _I}}}} \right),
\end{align}
and substituting the Laplace transform of interference for the typical user into the coverage probability expression, the theorem is proved.
\end{IEEEproof}
\end{theorem}

\begin{corollary}\label{alpha2}
\emph{Consider the special case when $\alpha_t = 2$, the closed-form coverage probability expression can be derived as  }
\begin{align}\label{EQ_nihe}
{{\rm{P}}_{cov,t}} = \sum\limits_{k = 1}^a {{{\left( { - 1} \right)}^{k + 1}}\binom{a}{k}} \frac{{\pi {\lambda _b}}}{{\left( {{\Xi _1} + {\Xi _2}} \right)}}
\end{align}
\emph{which can be derived via Eq.[2.3.3.1] in \cite{table}.}
\end{corollary}

\begin{corollary}\label{alpha4}
\emph{Conditioned on $\alpha_t = 4$, we can derive the closed-form expression of coverage probability for the typical user as  }
\begin{align}\label{EQ_nihe}
{{\rm{P}}_{cov,t}}{ =} \sum\limits_{k = 1}^a {\frac{{\pi {\lambda _b}}}{{2{{\left( { - 1} \right)}^{k + 1}}}}\binom{a}{k}} \sqrt {\frac{\pi }{{{\Xi _2}}}} \exp \left( {\frac{{\Xi _1^2}}{{4{\Xi _2}}}} \right){\rm{erfc}}\left( {\frac{{{\Xi _1}}}{{2\sqrt {{\Xi _2}} }}} \right),
\end{align}
\emph{which can be proved by Eq.[2.3.15.4] in \cite{table}.}
\end{corollary}

\begin{theorem}\label{CPc}
\emph{Since the channel quality of the connected user is lower than the typical user, the signals of the connected user will be directly decoded to ensure the performance. Thus, the coverage probability of the connected user is derived as  }
\begin{align}\label{eqCPc}
{{\rm{P}}_{cov,c}} = \exp \left( { - {\Xi _3}r_c^{{\alpha _c}}} \right)\exp \left( { - {\Xi _4}r_c^2} \right),
\end{align}
\emph{where ${\Xi _4} = \pi {\lambda _b}\left( {{}_2{F_1}\left( { - \frac{2}{{{\alpha _c}}},1;1 - \frac{2}{{{\alpha _c}}}; - \frac{{{\eta _c}\gamma _{\rm{c}}^{th}}}{{{a_c} - \gamma _{\rm{c}}^{th}{a_t}}}} \right) - 1} \right)$, ${\Xi _3} = {{{\eta _c}\gamma _{\rm{c}}^{th}{\sigma ^2}} \mathord{\left/
 {\vphantom {{{\eta _c}\gamma _{\rm{c}}^{th}{\sigma ^2}} {\left( {\left( {{a_c} - \gamma _{\rm{c}}^{th}{a_t}} \right){P_t}{C_c}} \right)}}} \right.
 \kern-\nulldelimiterspace} {\left( {\left( {{a_c} - \gamma _{\rm{c}}^{th}{a_t}} \right){P_t}{C_c}} \right)}}$ and ${\eta _c} = 1$. }
\begin{IEEEproof}
See \textbf{Theorem \ref{CPt}}.
\end{IEEEproof}
\end{theorem}

\vspace{-0.2cm}
\section{Numerical Results}

We exploit the numerical results to validate analytical coverage probability for typical users (\textbf{Theorem \ref{CPt}}) and connected users (\textbf{Theorem \ref{CPc}}) as upper bounds. Without otherwise specification, we set the numerical coefficients as: the noise power as ${\sigma ^2} =  - 170 + 10\log \left( {f_c} \right) + {N_f}=-90 $ dB with the bandwidth $f_c$ as $10$ MHz and the noise figure $N_f$ as $10$ dB, transmit power of users $P_t$ as $\left[0,30\right]$ dBm, pass loss exponents as $\alpha_c=\alpha_t = 4$, density of BSs as $\lambda_b = 1/(300^2\pi)$, thresholds $\gamma_{SIC}^{th}=\gamma_{t}^{th}=\gamma_c^{th}=10^{-2}$, and power allocation coefficients $a_c = 0.6$ and $ a_t = 0.4$. The number of the RISs $n$ and power consumption coefficient $\beta$ is defined in the following paragraphs.
\begin{figure}[!htb]
\vspace{-0.2cm}
\centering
\includegraphics[width= 3in]{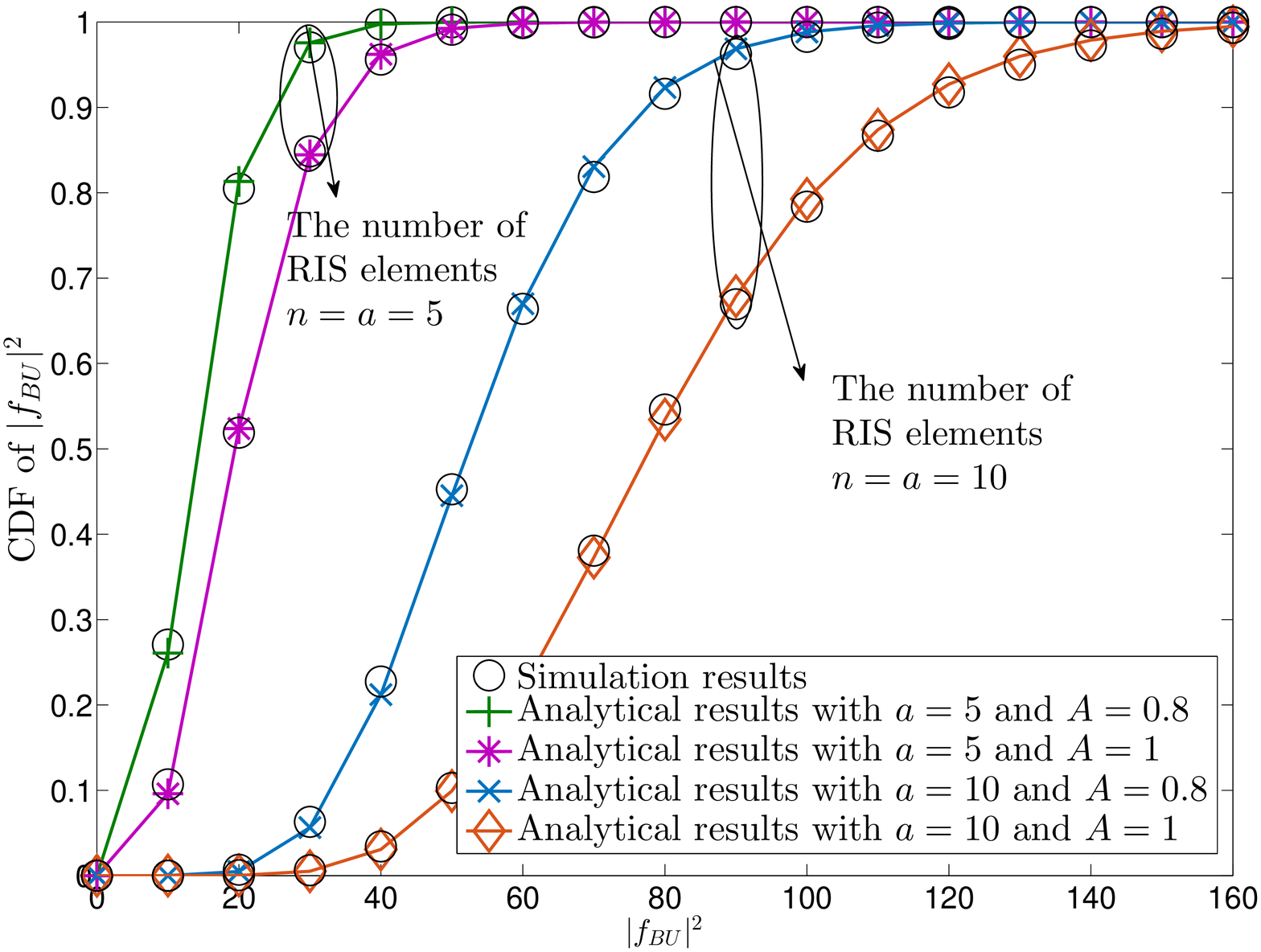}
\vspace{-0.4cm}
\caption{CDF versus the value of $\left|f_{BU}\right|^2$ with various numbers of RIS elements and power consumption coefficient.}
\label{figure2}
\vspace{-0.3cm}
\end{figure}
\begin{figure}[!htb]
\vspace{-0.2cm}
\centering
\includegraphics[width= 3in]{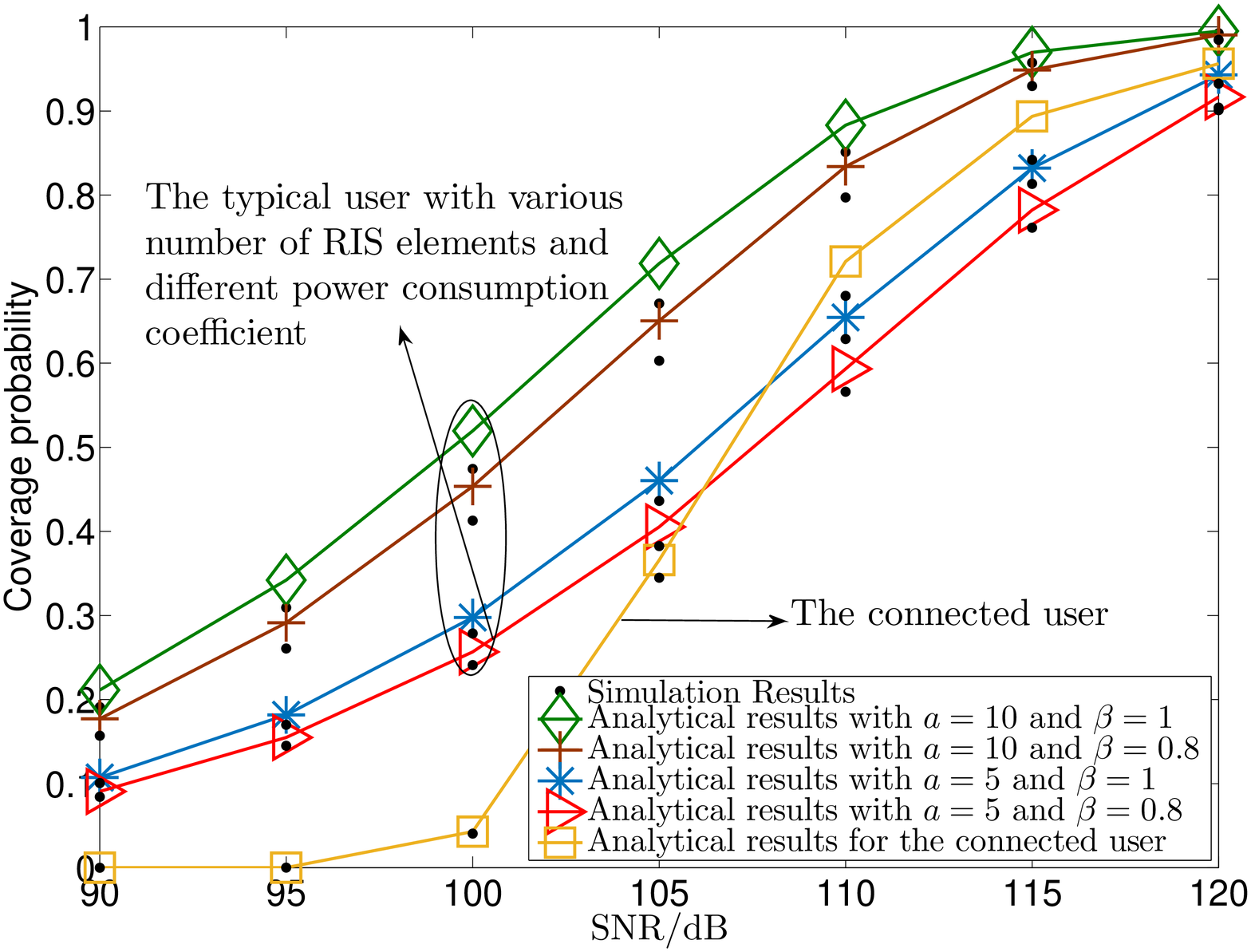}
\vspace{-0.4cm}
\caption{Coverage probability versus transmit SNR with various numbers of RIS elements $n=a=\{5,10\}$ and power consumption coefficient $\beta=\{1,0.8\}$.}
\label{figure3}
\vspace{-0.3cm}
\end{figure}
\begin{figure}[!htb]
\vspace{-0.2cm}
\centering
\includegraphics[width= 3in]{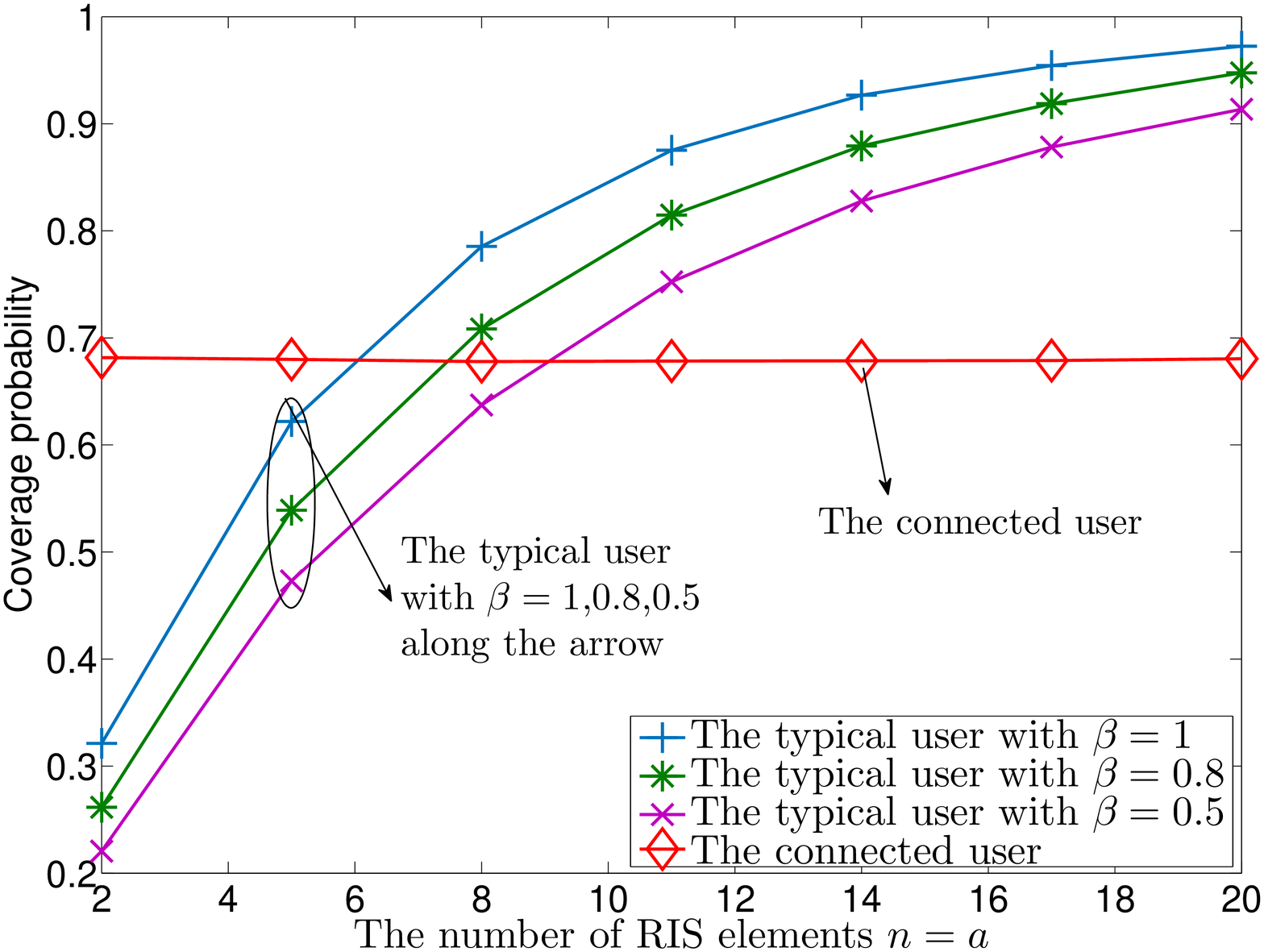}
\vspace{-0.4cm}
\caption{Coverage probability versus the number of RIS elements with $P_t = 20$ dBm and various power consumption coefficient $\beta=\{1,0.8\}$.}
\label{figure4}
\vspace{-0.6cm}
\end{figure}

Fig. \ref{figure2} illustrates that the validation of the RIS-aided channel model. One observation is that the number of RIS elements $n$ is the same as the shape coefficient of the Gamma distribution $a$. Another illustration is that the CDF of the RIS-aided channel model with a large number of RIS elements enhances more gently than the CDF with few RIS elements. Additionally, when we decrease the power consumption coefficient of the RISs $\beta$, the shape coefficient of the Gamma distribution $a$ is not influenced while the scale coefficient $b$ declines.

The coverage performance versus transmit SNR is investigated via Fig. \ref{figure3}, which demonstrates all the analytical results are upper bounds since we utilize a close upper bound assumption to derive the coverage probability. One observation is that increasing the number of RIS elements and power consumption coefficient can enhance the performance of the typical user. This is because enlarging the RIS element number and reducing the power consumption can improve the channel quality of the typical user.

Considering the number of RIS elements $n$, Fig. \ref{figure4} investigates the performance varied by $n$. Two observations are apparent to explain that: 1) the coverage probability with large numbers of RIS elements $n$ and power consumption coefficient $\beta$ outperforms that of low $n$ and $\beta$; 2) the RISs has no influence on the connected user since no RIS aids the connected user.

\vspace{-0.2cm}
\section{Conclusion}

This paper has investigated the coverage probability of a coherent scenario for RIS-aided multi-cell NOMA frameworks, where stochastic geometry models are exploited to capture the spatial effects of NOMA users. The RIS-aided channel model concluding the reflecting links and the direct links has been derived, which is closely modeled as a Gamma distribution. The analytical results have revealed that 1) the shape coefficient of the Gamma distribution $a$ is equal to the number of RIS elements $n$; and 2) the scale coefficient $b$ is equal to $n$ when the power consumption coefficient $\beta=1$ holds. Based on the channel model, the closed-form expressions of coverage probabilities have been derived. Numerical results has shown the conclusions that: 1) RISs can enhance the channel quality of its aided user; 2) the performance can be further enhanced via increasing the number of RIS elements $n$ and the power consumption coefficient $\beta$; and 3) RISs can provide high flexibility on decoding orders via adjusting the number of elements $n$.

\vspace{-0.2cm}
\section*{Appendix~A: Proof of Theorem~\ref{Channel1}} \label{Appendix:A}
\renewcommand{\theequation}{A.\arabic{equation}}
\setcounter{equation}{0}

When the number of RIS elements $n$ is large enough, we can assume the channel fading model as
\begin{align}
{\left| {{g_{BU}}} \right|^2} &= {C_t}d_t^{ - {\alpha _t}}{\left( {\sum\limits_{i = 1}^n {A{\beta _i}{c_{BR,i}}}  + {c_{BU}}} \right)^2} \notag\\
&\approx {C_t}d_t^{ - {\alpha _t}}{\left( {A\beta {S_K}} \right)^2},
\end{align}
where we set ${c_n} = {c_{BU}}$, ${c_k} = {c_{BR,k}}$ and ${S_K} = \sum\limits_{k = 1}^K {{c_k}} $ for $K=n+1$.

Firstly, since the $K$ variables have the same density distributions, we can briefly derive the distribution of the sum of $K$ independent random variables, denoted as ${S_K} $. Noted that a property for the distribution of ${S_K}$ is expressed as
\begin{align}
{{\cal L}_{{S_K}}}\left[ {{f_{{S_K}}}\left( x \right)} \right] = {\left\{ {{{\cal L}_{{S_K}}}\left[ {{f_{Rayleigh}}\left( x \right)} \right]} \right\}^K},
\end{align}
where ${f_{Rayleigh}}\left( x \right)$ is the PDF of Rayleigh fading channels. Hence, according to Eq. [2.3.15.1] in table \cite{table}, the Laplace transform of ${S_K}$ can be calculated as
\begin{align}
{{\cal L}_{{S_K}}}\left[ {{f_{{S_K}}}\left( x \right)} \right] = {\left( {\Gamma \left( 2 \right)\exp \left( {\frac{{{s^2}}}{8}} \right){D_{ - 2}}\left( {\frac{s}{{\sqrt 2 }}} \right)} \right)^K},
\end{align}
where ${D_{ - V}}\left( \cdot \right)$ is the parabolic-cylinder function shown in the index of notions in table \cite{table}.

Additionally, a special case with special values for the parabolic-cylinder function is expressed as ${D_{ - 2}}\left( {\frac{s}{{\sqrt 2 }}} \right) = \frac{1}{2}\exp \left( { - \frac{{{s^2}}}{8}} \right)\Psi \left( {1,\frac{1}{2};\frac{{{s^2}}}{4}} \right)$. Hence, the PDF of ${{S_K}}$ is expressed as
\begin{align}
{f_{{S_K}}}\left( x \right) = {\cal L}_{{S_k}}^{ - 1}\left\{ {{{\left( {\frac{1}{2}\Psi \left( {1,\frac{1}{2};\frac{{{s^2}}}{4}} \right)} \right)}^K}} \right\}\left( x \right).
\end{align}

When variable $x$ with its PDF $f_x(x)$, the PDF for $y = ax^2$ can be derived as $f_y(y)=\frac{1}{2a\sqrt(y/a)}\left[f_x(\sqrt(y/a))+f_x(-\sqrt(y/a))\right],y>0$. Hence, with the aid of the mentioned equation, the PDF of the equivalent channel model in power domain ${\left| {{g_{BU}}} \right|^2}=\Lambda  S_K^2$ can be obtained.

\vspace{-0.2cm}
\section*{Appendix~B: Proof of Lemma~\ref{interference1}} \label{Appendix:B}
\renewcommand{\theequation}{B.\arabic{equation}}
\setcounter{equation}{0}

Based on the Campbell theorem, the Laplace transform expression of the interference for the typical user is derived as
\begin{align}
{\cal L}_{(s)}^t {=} \underbrace {{\rm{E}}\left[ {\prod\limits_{{{\bf{x}}_I} \in {\Phi _r} \setminus {{\bf{x}}_B}} {\exp } \left( { - {\rho _I}s{P_t}{C_t}d_I^{ - {\alpha _t}}{{\left| {{f_{BU}}} \right|}^2}} \right)} \right]}_{{I_3}}\notag\\
 \times \underbrace {{\rm{E}}\left[ {\prod\limits_{{{\bf{x}}_I} \in {\Phi _r} \setminus {{\bf{x}}_B}} {\exp } \left( { - s\left( {1 - {\rho _I}} \right){P_t}{{\left| {{h_{BU}}} \right|}^2}{C_t}d_I^{ - {\alpha _t}}} \right)} \right]}_{{I_2}}.
\end{align}

With the aid of probability generating functional (PGFL) and $\int_A^\infty  {\left( {1 - \frac{1}{{{{\left( {1 + s{y^{ - \alpha }}} \right)}^N}}}} \right)ydy = \frac{{{A^2}}}{2}\left( {{}_2{F_1}\left( { - \frac{2}{\alpha },N;1 - \frac{2}{\alpha };} \right.\left. {-\frac{{  s}}{{{A^\alpha }}}} \right)} \right.} $ $\left. { - 1} \right)$, $I_2$ and $I_3$ can be derived as
\begin{align}
&I_2 {=} \exp \left( { - \pi {\lambda _b}d_t^2\left( {{}_2{F_1}\left( { - \frac{2}{{{\alpha _t}}},1;1 - \frac{2}{{{\alpha _t}}}; - {\xi _{\rm{1}}}s} \right) - 1} \right)} \right),\\
&I_3 {=} \exp \left( { - \pi {\lambda _b}d_t^2\left( {{}_2{F_1}\left( { - \frac{2}{{{\alpha _t}}},a;1 - \frac{2}{{{\alpha _t}}}; - {\xi _2}s} \right) - 1} \right)} \right).
\end{align}

\vspace{-0.2cm}
\bibliographystyle{IEEEtran}
\bibliography{mybib}
\end{document}